\documentclass[10pt,a4paper,final]{iopart}
\pdfoutput=1
\usepackage[utf8]{inputenc}
\usepackage[english]{babel}

\expandafter\let\csname equation*\endcsname\relax
\expandafter\let\csname endequation*\endcsname\relax

\usepackage{geometry}
\usepackage{amsmath, amssymb, amsthm}
\usepackage{tensor}
\usepackage{braket}
\usepackage{cases}
\usepackage{dsfont}
\usepackage[dvipsnames]{xcolor}
\usepackage{hyperref}
\hypersetup{
    colorlinks = true,
    allcolors = RoyalPurple
}
\usepackage{graphicx}
\usepackage{BOONDOX-cal} % more calligraphic letters, in particular H
\usepackage{tensor}
\usepackage{cite}
\usepackage{float}
\usepackage{placeins}
\usepackage[labelfont=bf]{caption}
\usepackage{subcaption}
\usepackage{hyperref}
\usepackage[normalem]{ulem}

\usepackage{tikz}
\usepackage{calculator}

\newcommand{\Uvar}{ \ensuremath{U_{\text{var}}}}
\newcommand{\Utrot}{\ensuremath{U_{\text{trot}}}}

\newcommand{\Uvaro}{ \ensuremath{U_{\text{var}, o}}}

\newcommand{\Uex}{\ensuremath{U_{\text{ex}}}}

\newcommand{\rema}[1]{#1}
\definecolor{swcolor}{rgb}{0.18, 0.70, 0.18}

\newcommand{\model}[0]{Transverse Field Ising }
\newcommand{\ham}[0]{Transverse Field Ising Hamiltonian }

\newcommand{\fadedline}[1]{
    \SUBTRACT{0.5}{#1}{\sub}
    \ADD{0.5}{#1}{\ad}
    \begin{center}
    \begin{tikzpicture}[]
        \fill[white, left color=white, right color=white, middle color=black] ( \sub \textwidth,0)--(0.5\textwidth, 0.1) (\ad\textwidth,0)--(0.5\textwidth, -0.1)--(\sub\textwidth,0)--cycle;
    \end{tikzpicture}
    \end{center}
}

\begin{document}
%\title[Classical Variational Optimization of Gate Sequences for Time Evolution]{Classical Variational Optimization of Gate Sequences for Time Evolution of Translational Invariant Quantum Systems}
\title{Variational Hamiltonian Simulation for Translational Invariant Systems via Classical Pre-Processing}

% Author order
\author{Refik Mansuroglu$^1$, Timo Eckstein$^{1, 2}$, Ludwig Nützel$^1$, Samuel A. Wilkinson$^1$, Michael J. Hartmann$^{1, 2}$}
\ead{\href{mailto:Refik.Mansuroglu@fau.de}{Refik.Mansuroglu@fau.de}, \href{mailto:Timo.Eckstein@fau.de}{Timo.Eckstein@fau.de}}
\address{$^1$Department of Physics, Friedrich-Alexander-Universität Erlangen-Nürnberg (FAU), Staudtstraße 7, 91058 Erlangen}
\address{$^2$Max Planck Institute for the Science of Light, Staudtstraße 2, 91058 Erlangen, Germany}

%\date{2021-05-31}

\begin{abstract}
\noindent The simulation of time evolution of large quantum systems is a classically challenging and in general intractable task, making it a promising application for quantum computation. A Trotter-Suzuki approximation yields an implementation thereof, where a higher approximation accuracy can be traded for an increased gate count. In this work, we introduce a variational algorithm which uses solutions of classical optimizations to predict efficient quantum circuits for time evolution of translationally invariant quantum systems. Our strategy can improve upon the Trotter-Suzuki accuracy by several orders of magnitude. It translates into a reduction in gate count and hence gain in overall fidelity at the same algorithmic accuracy. This is important in NISQ-applications where the fidelity of the output state decays exponentially with the number of gates. The performance advantage of our classical assisted strategy can be extended to open boundaries with translational symmetry in the bulk. We can extrapolate our method to beyond classically simulatable system sizes, maintaining its total fidelity advantage over a Trotter-Suzuki approximation making it an interesting candidate for beyond classical time evolution.
\end{abstract}


\maketitle

\section{Introduction}
Quantum simulation of real time evolution represents a natural application of quantum computation. In particular, non-equilibrium dynamics is known to lead to highly entangled states \cite{Trotzky_2012, Eisert_2015} and is therefore a strong candidate for quantum advantage, i.e. an application where quantum computers can compute quantities of interest that are not accessible to classical architectures \cite{Arute_19}. Given a time-independent Hamiltonian $H$ of a closed, physical system of $N$ qubits, one can define the time evolution operator as the unitary
\begin{align}
    \Uex = \exp{ \left( -i H \tau \right) },
    \label{eq:time_evol}
\end{align}
where here and throughout the paper we set $\hbar=1$. This operator propagates a quantum state $\ket{\psi(\tau_0)}$ along a time interval $\tau$ to $\ket{\psi(\tau_0 + \tau)}$. The dimension of the underlying Hilbert space grows exponentially in $N$, which makes simulation of large quantum systems intractable for classical computation, in general. A quantum computer, however, can directly operate on the $N$ qubits with quantum gates. When $H$ contains several non-commuting terms, we are unlikely to have a quantum gate of the form of (\ref{eq:time_evol}) at hand. Thus, a decomposition into experimentally accessible quantum gates is needed. One such decomposition can be achieved by a Trotter-Suzuki, or product formula, approximation \cite{Suzuki_85, Suzuki_91, Smith_2019, Kivlichan2020improvedfault}.

Besides product formula methods, there have been other approaches to implement algorithms for Hamiltonian simulation, such as by Quantum Signal Processing \cite{Low17}, Taylor series truncation \cite{Berry15, Childs18} and linear combination of unitaries \cite{Childs12}. Recently, the gate complexity in Trotter-Suzuki algorithms has been shown to scale comparably or even better than these so called post-Trotter methods \cite{Childs21}.

It has been shown \cite{Berry06} that for a permissible Trotter error $\epsilon$ of a $q^{th}$ order product formula, the gate count $G$ is bounded by
\begin{align}
    G \leq N 5^{2q} \frac{(NT)^{1+1/2q}}{\epsilon^{2q}}.
    \label{eq:gate_comp}
\end{align}
The Trotter error of $q^{th}$ order with Trotter number $m$ scales as $\epsilon = \mathcal{O}(\frac{T^{2q+1}}{m^{2q}})$. For a nearest-neighbor Hamiltonian, for instance, this bound can be further reduced to scale almost linear in system size and simulation time in the worst case \cite{Childs21}
\begin{align}
    G = \mathcal{O}\left( (N T)^{1+ \mathcal{O}\left( \frac{1}{q} \right) } \right).
    \label{eq:gate_compII}
\end{align}
For the simulation of Hamiltonians with local interactions \cite{Haah21} the gate complexity of higher order Trotter-Suzuki formulas meets a lower bound (for $q \to \infty$ in (\ref{eq:gate_comp})). This lower bound is realised with product formulas \cite{Childs21} as well as decompositions using Lieb-Robinson bounds \cite{Haah21}.

Since quantum simulation on today's Noisy Intermediate Scale Quantum (NISQ) devices \cite{Preskill2018quantumcomputingin} suffers from gate imperfections, variational methods have been put forward to reduce the total gate count and circuit depth \cite{rattew2020domainagnostic, Li_17, Crstoiu_2020, McClean_2016}. While a Trotter-Suzuki decomposition already yields a good approximation to a Hamiltonian-generated unitary time evolution, it has been shown that alternative methods can improve the total fidelity in specific cases by trading off the improvement of approximation accuracy against the gate count. These alternative methods include randomization \cite{Campbell17, Campbell19, Ouyang2020compilation, faehrmann2021, Childs2019fasterquantum, Chen2021}, tensor network techniques \cite{Barratt2021}, variational methods \cite{Jones_19, Barison21, Lubasch21, lau2021nisq, gibbs2021longtime, commeau2020variational, heya2019subspace} and also a non-variational optimization of product formulas if there are only few non-commuting terms in the Hamiltonian \cite{BARTHEL2020168165}. While the optimization potential on the asymptotic behavior of the gate complexity for the simulation of local interaction Hamiltonians is moderate, constant factor improvements (in $N$) can already result in quantum advantage under NISQ conditions. This is due to the fact that the fidelity for a desired output state decreases exponentially in the gate count for finite fidelity gates.

Apart from the dependence on $N$ and $T$ as in (\ref{eq:gate_comp}), there is a constant factor $5^{2q}$ which grows with the order of the Trotter-Suzuki formula $q$ \cite{Berry06}. This factor makes higher order product formulas unfavorable for NISQ devices. On the other hand, finding an optimal gate sequence with variational methods can improve on that factor by keeping $q$ low and thus represent a good NISQ solution where simulation times and system sizes are moderate such that asymptotic behavior is not dominating the gate number. As product formulas represent a universal, i.e. model-independent, method for Hamiltonian simulation, it is not surprising that one can optimize it for a specific Hamiltonian.

Variational methods make use of this optimization potential in different ways. They might exploit hardware-efficiency for a specific ansatz state \cite{Lubasch21, Barison21, lau2021nisq}, propose an efficient diagonalization of the Hamiltonian within the available gate-set \cite{gibbs2021longtime, commeau2020variational} or focus on low-energy initial states \cite{heya2019subspace}. 

In this work, we tackle two difficulties that arise in these variational methods. First, employing a quantum optimization routine requires multiple iterations with high sampling cost for each. This is a manifest problem in NISQ algorithms \cite{harrigan2021, Bharti22, Cerezo_2021, Tilly21}. Our method allows to move the optimization to a classical pre-processor for translationally invariant Hamiltonians. Second, most of recent variational methods depend on a specific initial state for which the dynamics are calculated \cite{Barison21, Lubasch21, lau2021nisq, gibbs2021longtime}. In contrast, a time-universal simulation algorithm needs to be applicable for any initial state. In particular, the same circuit can be applied multiple times on a desired initial state and all the intermediate states. The method we present is state agnostic and therefore not limited to applications on states with bound entanglement as various classical methods \cite{Or_s_2019}.

\section{Main Results}
We present an initial state agnostic variational algorithm for the simulation of translational invariant Hamiltonians where the heavy sampling effort of the optimization is avoided via classical pre-processing.

For a translationally invariant Hamiltonian, our generalization scheme allows to obtain the quantum circuit of a large system from the optimal solution of a small, classically solvable system. The upscaling to larger systems can be performed qubit-by-qubit if the variational parameters are chosen to be translationally invariant.

\begin{figure}
    \centering \includegraphics[width=0.975\textwidth]{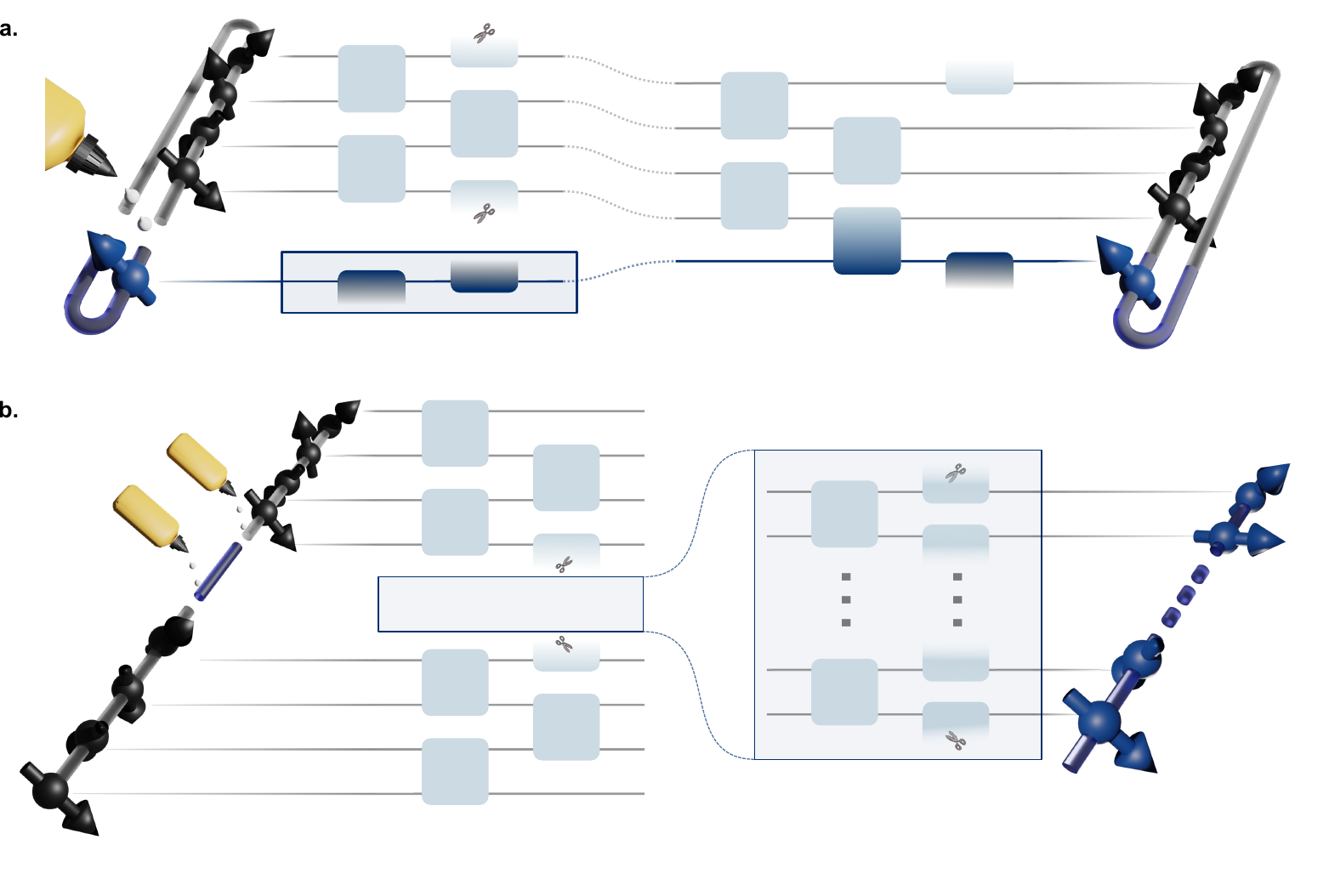}        
    \caption{\textbf{a. Upscaling procedure for a one dimensional system with periodic boundaries.} The classical variational optimization is conducted on the gray small system (here 4 qubits). To extend this small system to a larger system qubit-by-qubit, for each interaction layer, the variational parameter of the small system's two-qubit gates is reused. To include the new qubit (blue), the small periodic system is cut at one point and two-qubit gates connecting the open points of the cut system and a new qubit are added (compare also circuit picture in the middle). This extension strategy is motivated by the translational invariance of the system.
    \textbf{b. Upscaling procedure for a one dimensional system with open boundaries.} For open boundaries the translational invariance that was used previously as motivation of the upscaling strategy is no longer fulfilled. However, in this case two small systems, one with open and one with periodic boundaries, can be used. The open boundary system (gray) is cut in half and in its middle an arbitrary number of qubits using the optimal variational parameters of the periodic boundary system (blue) are inserted applying the same two-qubit gate gluing procedure as before. This algorithm is illustrated in the middle in the quantum circuit picture.}
    \label{fig:chain}
\end{figure}

For open boundary conditions, which break translational invariance, we propose a variant of the method which uses two classical optimization routines, one which optimizes the open boundary system and one which yields optimal parameters for a periodic boundary system. The upscaled system then combines the first set of parameters for the qubits near the boundary and the second set for qubits within the bulk. Two dimensional upscaled open boundary systems potentially require classical optimization of more than one small open boundary system, but otherwise follow the same procedure. This scenario is favored by modern NISQ devices which would require a number of SWAP operations to implement periodic boundaries on two dimensional architectures. Also, open boundary systems are expected to pose larger challenges for classical simulation methods since no translation symmetry can be exploited in classical algorithms. The process of gluing together optimized systems is discussed in detail in section \ref{sec:glue} and illustrated in Figure \ref{fig:chain}.

The presented methods are numerically exemplified on a one- and two-dimensional \model Model. On small systems for which we can compare the quantum circuit with exact time evolution, our variational approach improves upon the Trotter-Suzuki decomposition in terms of accuracy by up to 3 orders of magnitude in the cost function and at the same time reduces the gate count by a factor of 2. We are then able to suggest quantum simulation circuits for larger systems using the optimal parameters from classical pre-processing, which outperform the Trotter-Suzuki ansatz in a similar way.

Under NISQ conditions, i.e. with imperfect gate fidelities, such improvements turn out to be decisive as to whether an algorithm can be executed with a Trotter-Suzuki ansatz. These scenarios appear when the simulation time $\tau$ for fixed circuit depth is large, see Figures \ref{fig:Vary_N} and \ref{fig:NISQ}.

In numerical experiments with the one-dimensional \model Model, we found for example that with 50 qubits, a single gate error of $0.1 \%$ and $\tau = 2.5$ our algorithm obtains a NISQ preparation infidelity of ca. $29 \%$, being only a $7 \%$ increase from the minimal infidelity due to gate errors. Compared to approx. $57 \%$ infidelity of the best Trotter-Suzuki ansatz. For this numerical experiment, the optimal parameters for the time evolution on a 6 qubit chain have been used.

\section{Variational Optimization}
\label{sec:Opt}
We consider a $p$-local quantum many-body Hamiltonian $H$ which consists of $N \cdot A$ terms
\begin{align}
    H = \sum_{n=1}^N \sum_{a = 1}^A c_a H_{n, a},
    \label{eq:Ham}
\end{align}
where $a \in \{1, ..., A\}$ counts different interaction types and $n\in \{1, ..., N\}$ denotes the first qubit on which $H_{n, a}$ is acting. For a $k-$local term ($k \leq p$), for instance $H_{n, a}$ acts on qubits $n, ..., n+k-1$ modulo periodic boundary conditions.

The $\{H_{n, a}\}_{(n, a)}$ of the Hamiltonian are chosen such that each $H_{n, a}$ is the generator of a quantum gate that can straight-forwardly be implemented on the considered hardware (depending on the $p$-locality of $H$ and on the quantum architecture, further decompositions into experimentally accessible gates might be involved).

The unitary time evolution generated by $H$ can be approximated by a first-order Trotter decomposition \cite{Suzuki_85}
\begin{align}
    \Uex &= \exp{\left( -i \tau \sum_{n, a} c_a H_{n, a} \right)}  = \left[ \prod_{n, a}^\leftarrow \exp{\left( \frac{- i \tau}{m} c_a H_{n, a} \right)} \right]^m + \mathcal{O}\left( \frac{\tau^2}{m} \right),
    \label{eq:Trotter1}
\end{align}
where $m$ denotes the Trotter number and the arrow denotes an ordering which starts with the lowest pair $(n, a)$ on the right. The ordering inferred on the tuple $(n, a)$ is not unique and only needs to be fixed when constructing the corresponding circuit. The leading order error of this approximation is given by commutators between the Hamiltonian terms \cite{Childs21}. A $q^\text{th}$ order Trotter-Suzuki decomposition while using more gates improves the error to leading order $\mathcal{O}(\frac{\tau^{q+1}}{m^{q}})$ \cite{Suzuki_85, Suzuki_91}. We will stick to first order product formulas as a benchmark in the main text of this paper, since higher order approaches use exponentially more gates in $q$ and hence yield worse total fidelities compared to increasing the Trotter number $m$. We show in \ref{app:2OSuzuki} that also smaller Trotter numbers for higher order Trotter sequences do not save enough gates to yield a better simulation. 

Our method is independent of the particular variational quantum circuit ansatz $\Uvar$, which is a gate sequence of parametrized unitaries
\begin{align}
    \Uvar(\vec{\theta}) = U_{1} (\theta_{1}) \cdot U_{2} (\theta_{2}) \cdots
\end{align}
We choose the first order Trotter-Suzuki approximation from (\ref{eq:Trotter1}) as a starting point for the circuit ansatz, as this ensures an accuracy that is at least as good as Trotter-Suzuki approximation. In order to exhaust the potential of the ansatz from (\ref{eq:Trotter1}), we will substitute the coefficients $\frac{\tau}{m} c_a$ for each of the Hamiltonians $H_{n, a}$ at each of the $m$ sequences by a variational parameter $\theta_{r, a}$ with $r \in \{1, ..., m\}$ labeling the layers and $a \in \{ 1, ..., A \}$ the different gates. The resulting ansatz reads
\begin{align}
    \Uvar(\vec{\theta}) = \prod_{r = 1}^m \left[ \prod_{n, a}^\leftarrow \exp{\left( -\frac{i}{2} \theta_{r, a} H_{n, a} \right)} \right].
    \label{eq:Var_seq}
\end{align}
Similar to the Trotter-Suzuki approximation, the variational ansatz admits an approximation order $m$ which controls the number of parameters. There are thus $A \cdot m$ different parameters to tune. Similar ansätze as in Eq. (\ref{eq:Var_seq}) are known as variational Hamiltonian ansatz \cite{Cerezo_2021} for adiabatic ground state preparation, where they are used with a different cost function than in our case.

The optimization objective is to determine a set of optimal parameters $\theta_{r,a}$ for which the exact time evolution $\Uex$ is approximated best by $\Uvar$. The optimal parameters are found by minimizing a cost function measuring the distance of unitaries. We choose a measure of distance induced by the Frobenius norm
\begin{align}
    C(U_\text{var}, \vec{\theta}) &= \frac{1}{2^N}  ||\Uvar(\vec{\theta}) - \Uex||_F^2 = \frac{1}{2^N} \sum_{a, b=1}^{2^N} \lvert \Uvar \tensor{}{^a_b} - \Uex \tensor{}{^a_b} \rvert^2  \quad \in [0, 1],
    \label{eq:Frob_cost}
\end{align}
where $\tensor{U}{^a_b}$ denotes the elements of a matrix representation of the operator $U$. The additional factor $\frac{1}{2^N}$ eliminates the scaling of $C(\vec \theta)$ in system size $N$. The exact operator of time evolution $\Uex$ serves as a benchmark which is agnostic of an initial state. In a similar fashion, we can consider $C(U_\text{trot}, \theta)$ as a measure of distance between the Trotter-Suzuki ansatz and exact time evolution. 

The cost function (\ref{eq:Frob_cost}) is a measure of how well time evolution is approximated on average in the full Hilbert space. As we are considering finite-dimensional Hilbert spaces, any error of observable dynamics can be upper bounded by a distance measure
\begin{align}
    | \braket{\psi|\Uex^{\dagger} \, \mathcal{O} \, \Uex| \psi} - \braket{\psi|\Uvar^{\dagger}(\theta) \, \mathcal{O} \, \Uvar(\theta)| \psi} |  \leq 2 ||\mathcal{O}||_\infty \, ||\Uvar(\theta) - \Uex||_\infty = 2\kappa ||\mathcal{O}||_\infty \, ||\Uvar(\theta) - \Uex||_F,
    \label{eq:upper_bound}
\end{align}
where $||.||_\infty$ denotes the operator norm. The constant $\kappa$, in general, depends on the distribution of the error throughout the Hilbert space. For a concentrated error, we have $\kappa = 1$ while for the pathological case that only one state is simulated badly while every other state has perfect fidelity, the upper bound gets loose by a factor of $\kappa = 2^n$. For the Trotter algorithm it is known that most of the time, the errors will be concentrated \cite{Chen2021} making the cost function \eqref{eq:Frob_cost} a good estimate for the performance of the algorithm.

As an example, the optimization of one- and two-dimensional \model Models on rectangular lattices is discussed in section \ref{sec:numerics}, but for now let us assume, we have found a set of parameters such that $C(\vec{\theta})$ as defined in (\ref{eq:Frob_cost}) is minimal. Note that (\ref{eq:Frob_cost}) involves the full time evolution operator which is not accessible on a quantum computer, but has to be calculated on a classical computer. Consequently, there is no quantum advantage, if $\Uvar$ is deployed to a quantum computer. However, a very meaningful, NISQ-friendly application arises, when the found optimal parameters of the trained smaller system are applied to a larger unseen system, which is explained in the following.

\section{Upscaling System Size}
\label{sec:glue}
Hamiltonians of many-body quantum systems often enjoy translational symmetries, such that larger systems look similar to smaller systems, locally. In particular, $H_{n, a}$ of (\ref{eq:Ham}) is the same operator for all $n$ except that it acts on different qubits. Our method exploits this symmetry and generalizes solutions of variational optimizations to larger quantum systems. 

To begin with, the upscaling algorithm is explained for fully translational invariant systems. These necessarily have periodic boundaries. Since some modern quantum hardware, however, naturally comes with open boundary conditions for the implementable gates which disrupt translational invariance, we also extend the method by systems, where the translational symmetry is only broken by open boundaries. 

\subsection{Translationally Invariant Systems}
Translationally symmetric systems can be scaled up by successive application of the gates from the small system to the new qubits. Figure \hyperref[fig:chain]{\ref*{fig:chain} a}, for instance, shows a one-dimensional spin chain with periodic boundary conditions. For nearest-neighbor interactions, the interaction between the first two qubits is the same as the interaction between any other neighboring pair of qubits. From this translational invariance, ansätze for larger quantum systems are generated by copying the quantum gates of the small circuit. More precisely, we apply the quantum gates, which acted on qubits $N$ and 1 before, after extension to qubits $N$ and $N+1$ as well as 1 and $N +1$, respectively. In principle, this qubit-by-qubit upscaling scheme can be applied repetitively to extend an initially few qubit time evolution circuit to a many qubit one. The final extended variational circuit can then be easily inferred from (\ref{eq:Var_seq}) to be
\begin{align}
    \Uvar(\vec{\theta}) = \prod_{r = 1}^m \left[ \prod_{\substack{n=1, ..., N+K \\ a=1, ..., A}}^\leftarrow \exp{\left( -\frac{i}{2} \theta_{r, a} H_{n, a} \right)} \right].
    \label{eq:periodic_gluing}
\end{align}
where $K$ denotes the total number of added qubits. Note that (\ref{eq:periodic_gluing}) encodes more than just tensorating copies of the systems, but restructures gates such that entanglement between the former $N$ qubits and the new $K$ qubits is created (cf. Figure \hyperref[fig:chain]{\ref*{fig:chain} a}).

Very importantly, the method itself is insensitive of the precise model as long as it is translational invariant. As before, we will stick to the Trotter-Suzuki inspired variational ansatz and only need to define the Hamiltonian of the extended system $H^{(N+1)}$ and then use the gate sequence from the copied addends in the Hamiltonian as before. To get this, we subtract the old boundary term and insert the new terms for the new qubit 
\begin{align}
    H^{(N+1)} = H^{(N)} &- \sum_a \sum_{k=1}^{p-1} H_{N-(k-1) \rightarrow p-k, a}  + \sum_a \sum_{k=1}^{p} H_{N-(k-2) \rightarrow p-k, a} 
    \label{eq:glued_ham}
\end{align}
where we used the notation $H_{n_1 \rightarrow n_2, a}$ to emphasize that this term acts on the qubits $n_1, n_1 + 1, ..., n_2$. This procedure is also sketched for $p=2$ in Figure \hyperref[fig:chain]{\ref*{fig:chain} a)}. Ultimately, this upscaling can be repeated to end up with the Hamiltonian for $N+K$ qubits. 

This process might seem redundant at first. However, if we recall the variational method to improve on trotterized time evolution, we have just derived a generalization scheme which allows us to use the parameters $\vec{\theta}$ found for the $N$ qubit system to suggest a decomposition of quantum gates for the simulation of the $N+K$ qubit system. This is done by, again, breaking the periodic boundary conditions and gluing the gates corresponding to the $K$ new qubits but now with the variational parameters instead of the Trotter parameters $\frac{\tau}{m} c_a$.

It becomes apparent that the gluing method is identical to Trotterization on a coarser level. This can be understood best, when looking at the reverse process using the Trotter-Suzuki parameters $\frac{\tau}{m} c_a$. Given the Hamiltonian $H^{(N+K)}$ of the large system, $\exp(-i\tau H^{(N+K)})$ is split into the product $N+K$ identical gates with the only difference of acting on different qubits as described in (\ref{eq:glued_ham}). Two subsequent Trotterizations, first into $H^{(N)}$ and $H^{(K)}$ and afterwards into 2-qubit gates yield the same result as a Trotterization directly into 2-qubit gates. 
As a result, if the variational parameters are reset to the Trotter-Suzuki parameters, there is no difference in a Trotter-Suzuki decomposition of $\exp(-i \tau H^{(N+K)})$ and gluing $K$ qubits to the system $H^{(N)}$. 

\rema{Since the leading error of Trotter approximations is given by commutators between Hamiltonian terms, this error scales linearly in the qubit number for local Hamiltonians \cite{Childs21}. Due to the similar structure of our ansatz in equation \eqref{eq:periodic_gluing}, one may expect comparable scaling for our variational method. This suggests that the upscaling procedure we employ only causes a degrading of approximation accuracy that scales linearly in the system size. This expectation is confirmed by our numerical experiments presented below.}

\subsection{Open Boundary Conditions}
\label{sec:open_bound}
On some quantum hardware, such as superconducting circuit devices, the implementation of periodic boundary conditions on two-dimensional lattice models requires additional SWAP gates. Hence, the simulation of physical systems with open boundary conditions is better suited for such devices, particularly under NISQ conditions where each additional operation exponentially decreases the success probability. To adapt the upscaling scheme to systems with open boundary conditions, additionally to a small system with periodic boundaries, a small system of size $N_o$ with open boundaries is variationally optimized. The corresponding Hamiltonian has the same form as (\ref{eq:Ham}) but without terms reaching beyond the periodic boundary. An intuition for why this should work is that if the depth of the circuit is short enough, the bulk qubits of the open system are affected by approximately the same operations as for a periodic boundary system. This is why we are allowed to glue an arbitrary periodic system into the middle of an open boundary system.

Time evolution can be approximated by the operators $\Utrot {}_{, o}$ and $\Uvar {}_{, o}(\vec \theta)$ in an analogue way as before. The upscaled system is constructed from both the small periodic and non-periodic one, by cutting the small open boundary system of size $N_o$ in half and inserting an arbitrary number $K$ of periodic boundary qubits in the middle (see Figure \hyperref[fig:chain]{\ref*{fig:chain} b}.). The gluing of the gates for each time step works as described before. The final circuit for time evolution of a single time step ($m=1$) for the glued system with $N_o + K$ qubits with open boundaries reads 
\begin{align}
    \Uvaro^{(1, N_o + K)} \approx \, &\Uvaro^{(1, \frac{N_o}{2})} (\vec \theta_{o}) \, \Uvar(\vec \theta) \, \Uvaro^{(K + \frac{N_o}{2} + 1, K+N_o)} (\vec \theta_{o})
    \label{eq:open_construct}
\end{align}
where we used $\Uvar(\theta)$ from (\ref{eq:periodic_gluing}) for the time evolution of the bulk. $\Uvaro^{(1, X)}$ denotes circuit simulating the open boundary dynamics on qubit 1 to $X$. Furthermore, we split the optimized parameter set $\theta_o$ of the open boundary $N_o$ qubit simulation in half in order to use them for the time evolution of the left (or respectively right) boundary. For the sake of simplicity, we assume $N_o$ to be even. This way, the three unitaries in (\ref{eq:open_construct}) correspond to the three segments of Figure \hyperref[fig:chain]{\ref*{fig:chain} b} that are being glued. The boundary parameters $\theta_o$ and the bulk parameters $\theta$ result from two separate optimizations. We are thus making an approximation neglecting open boundary effects on the bulk qubits. We will see in numerical experiments that for a one- and two-dimensional \model Model it suffices to optimize only few dedicated boundary gates separate from the periodic bulk.

\section{\model Model}
\label{sec:numerics}
To numerically test the performance of the variational gluing method in comparison to Trotter-Suzuki approaches, the \model Model (TFIM) in one and two dimensional is studied. For this, we simulate the exact time evolution $\Uex$ generated by the Hamiltonian
\begin{align}
    H_{TFIM} = J_z \sum_{\langle i, j \rangle} Z_i Z_j + h_x \sum_i X_i,
    \label{eq:2DHam}
\end{align}
with the Pauli operators $X, Z$ and the coupling constants $J_z$ and $h_x$. We denoted a Pauli $Z$ operator acting on the $i^\text{th}$ qubit by $Z_i$ (and analogously for $X$). Furthermore, the notation $\langle i, j \rangle$ only counts nearest neighbor terms and we implicitly understood periodic boundary conditions. The \ham relates to our previous notation via $H_{n, a=1} = Z_n Z_{n+1}, H_{n, a=2} = X_n$ and $c_1 = J_z, c_2 = h_x$. To demonstrate our method on another model, we show additional results for an $XY$ model with a transverse field in \ref{app:XY}.

The cost values for the Trotter-Suzuki $\Utrot$ and variational sequence $\Uvar$ are compared for different system sizes, as well as for multiple repetitions of the circuit (i.e. multiple time steps). To emphasize that near term NISQ hardware can benefit from our classical pre-training algorithm for time evolution, simulation results of 2-point spin correlations including gate infidelities are presented.

We simulated and optimized systems with a total number of up to 12 qubits in the discussed fashion. For the study of larger systems limitations of memory and time resources forced us to approximate the cost function (\ref{eq:Frob_cost}) by sampling over a set of random input vectors $\mathcal{V}$
\begin{align}
    C_{var}(\vec \theta) &\approx \frac{1}{|\mathcal{V}|} \sum_{\psi \in \mathcal{V}} \left( 1 -  |\braket{\psi| \Uex^\dagger \Uvar(\vec\theta) | \psi}| \right),
    \label{eq:Frob_cost_samp}
\end{align}
where the absolute value $|.|$ ensures the equality of the two operators up to a multiplicative global phase. This way, we are able to present results for system sizes up to 24 qubits. For the circuit simulations we used the python library Cirq \cite{cirq_developers_2021_4586899} and for optimization we used the \textit{Adam} gradient descent method \cite{Kingma2015AdamAM}. In all the optimizations, we have used a maximal number of traning steps of $10^4$, a learning rate of $10^{-4}$ and the hyperparameters $\beta_1 = 0.9, \beta_2 = 0.99$ to control the decay rates of the gradient and squared gradient average. 

There is an ambiguity in the choice of the ordering of the gate sequences which correspond to $Z_i Z_j$ and $X_i$. This leads to different operators when doing Trotterization. Here, we fixed the ordering for the sake of an efficient implementation to
\begin{align}
    \Uvar(\vec\theta) &= \prod_{r = 1}^m \Bigg[ \prod_{\substack{n = 1 \\ n \text{ odd}}}^{N} e^{-i \theta_{r, 1} Z_{n} Z_{n+1}}  \prod_{\substack{n = 1 \\ n \text{ even}}}^{N} e^{ - i \theta_{r, 1} Z_{n} Z_{n+1}} \prod_{n = 1}^N e^{ -i \theta_{r, 2} X_n } \Bigg].
    \label{eq:Uvar_model}
\end{align}
Each of these gates can be implemented directly on current superconducting hardware via tunable couplers \cite{Salathe15, Collodo20, Foxen20, Mahdi19, Lacroix20}. In this numeric example, four cases are studied. These are one- and two-dimensional systems with either periodic or open boundary conditions. In the case of two-dimensional periodic lattices, the gluing can be performed in either of the two dimensions by adding a row or column of qubits analogously to the discussion in section \ref{sec:glue}. Demonstration of scaling up in system size on two-dimensional systems with two open boundaries is computationally prohibitive, as the smallest fully open, two-dimensional and non-trivially glued system would be of size $5 \times 7$ (cf. \ref{app:2D_open}). We therefore employ open boundaries in only one direction, while retaining periodic boundary conditions for the orthogonal direction. Optimal parameters can be classically found but no measure of accuracy can be evaluated any longer as we exceed our limit of numerical feasibility with a total size of 35 qubits.

\subsection{Single Time Step}
\label{sec:performace}
\begin{figure}
    \centering
    \includegraphics[width=0.975\textwidth]{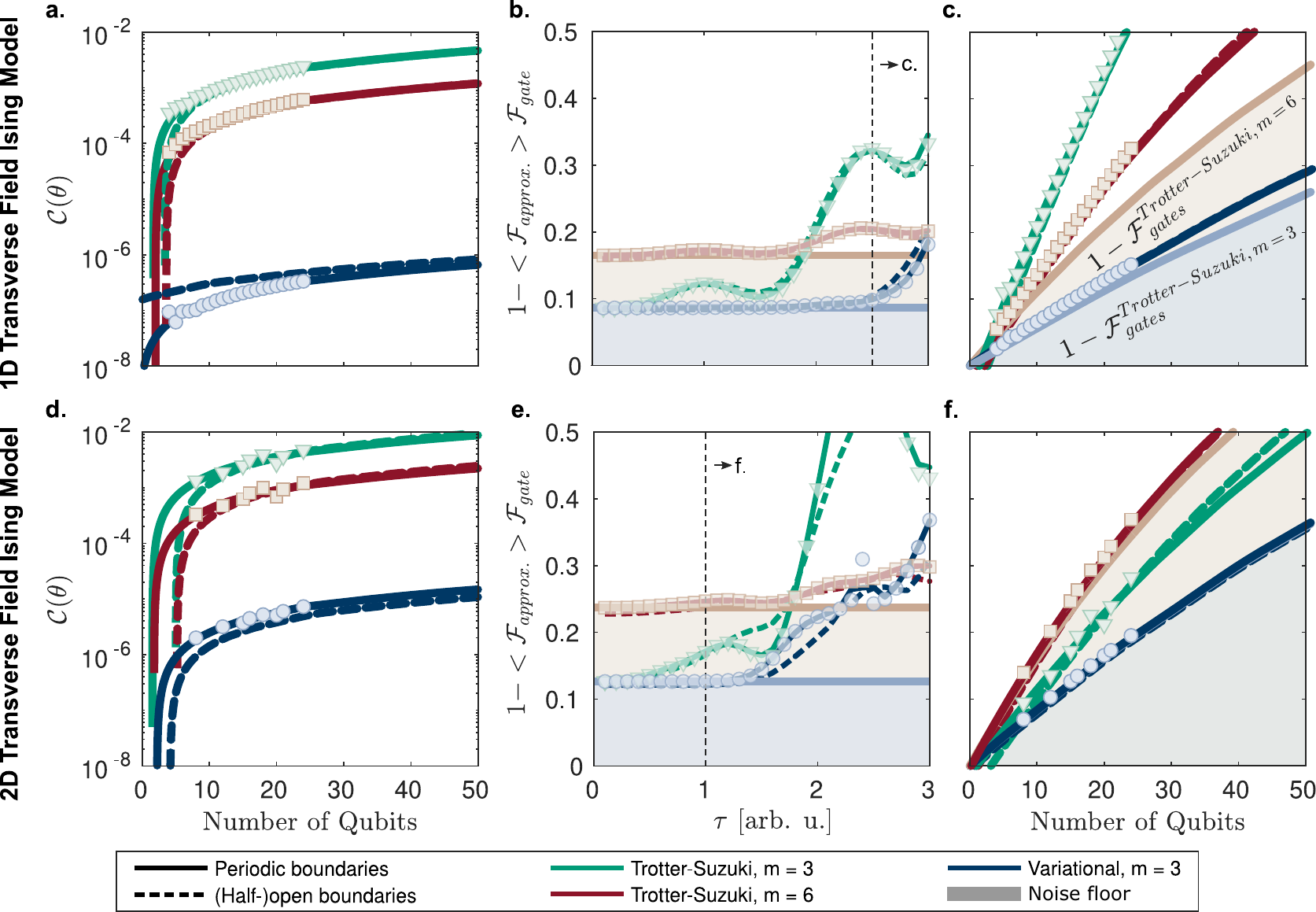}
    \caption{ \textbf{ System size upscaling of variational and Trotter sequence for a single time step $\tau$.} In the simulations of several system sizes and for one- and two-dimensional systems with either periodic or (half-)open boundaries, $J_z = 1.0$ and $h_x = 0.25$ were used. The data points show numerical evaluations of the cost function for the periodic boundary case. The solid and dashed lines in a. and d. represent a linear fit in the number of qubits for both periodic and (half-)open boundaries. \textbf{a, d.} Comparison of cost function for a time step of $\tau = 0.3$. The optimization was performed on 6 qubits for the one-dimensional and on a $3 \times 3$ grid for the two-dimensional systems. \textbf{b, e.} Variation of the single time step $\tau$ versus overall infidelity from algorithmic and gate error ($p_g = 99.9\%$) for $N = 15$ qubits ($3 \times 5$ for $d=2$). The noise floors generated by the sole gate infidelity (independent of $\tau$) are depicted by a shaded region in blue ($m=3$) and red ($m=6$). For small time steps, the gate infidelity dominates the error, while for large time steps, the algorithmic error takes over. For a one-dimensional system, this turning point lies around $\tau=2.5$ and $\tau = 1.5$ for two-dimensional systems. The turning points of the Trotter-Suzuki decompositions are earlier. \textbf{c, f.} Scaling of the total infidelity in system size (analogue to a. and d.) for $\tau = 2.5$ ($\tau = 1.0$ for $d=2$). Here, again, the optimal parameters found from a 6-qubit chain ($3 \times 3$ grid for $d=2$) are scaled up to a total size of 24 qubits and fitted as in a. and b..}
    \label{fig:Vary_N}
\end{figure}

To begin with, our variational circuit is compared to Trotter-Suzuki approaches with equal or higher gate count and depth, where higher gate count is usually traded against a better approximation. Afterwards, the optimal parameters are reused in larger systems and also compared to the Trotter-Suzuki benchmarks. Figures \hyperref[fig:Vary_N]{\ref*{fig:Vary_N} a} and \hyperref[fig:Vary_N]{d.} show that the variational ansatz is able to beat Trotterizations with a higher Trotter number along all numerically feasible system sizes (while being less interesting as it requires more gates, higher order Trotter-Suzuki approximations can also be beaten, see \ref{app:2OSuzuki}). Notably, the cost function values for upscaled systems hardly differ from the systems on which the optimization is performed. For the one-dimensional system, optimization was performed on a 6-qubit chain. An upscaling to 24 qubits only raises the cost value by a factor of 3. Analogously, for a two-dimensional system, the optimal parameters found for a $3 \times 3$ grid are extended to a $4\times 6$ grid by increasing the cost value by around the same factor. The cost values for the two-dimensional models are generally higher since each qubit has more interactions partners.

A linear fit in system size gives an estimate of how the improvement develops for system sizes which can no longer be classically simulated. In all studied cases, we observe an improvement factor in cost of the variational method compared to Trotter-Suzuki of more than 1000 in the one-dimensional and more than 100 in the two-dimensional case while using only half of the number of gates. The linear fits indicate that these improvement factors are constant over several larger system sizes. 

NISQ gates have finite infidelities, leading to large compound errors when gate counts are high. It is therefore important to be as gate efficient as possible when running Hamiltonian simulation on NISQ hardware. To take into account gate infidelities, we quantify the total infidelity of a simulated NISQ experiment by
\begin{align}
    1 - \mathcal{F}_{NISQ} = 1 - \braket{\mathcal{F}_{approx}} \mathcal{F}_{gates},
    \label{eq:cost_infid}
\end{align}
where $<\mathcal{F}_{approx}>$ denotes the averaged fidelity of the output state (variational or Trotter) when the hardware runs without error and $\mathcal{F}_{gates}$ is the average fidelity of the implementations of the employed gates. $<\mathcal{F}_{approx}>$ is inferred from the cost function used for optimization, where (\ref{eq:Frob_cost}) takes an average over the whole Hilbert space while (\ref{eq:Frob_cost_samp}) only takes an average over a randomly chosen subset. For the gate fidelity $\mathcal{F}_{gates}$, we consider a simple error model assuming the same fidelity $p_g$ for each individual gate. The circuit fidelity is given as the product of the individual gate fidelities for every gate in the circuit, i.e. $\mathcal{F}_{gates} = p_g^G$.

The total number of gates $G$, in general, is determined by the Trotter number $m$. In particular, for the Trotter-Suzuki decomposition of a \model Model with periodic boundaries, depth $D$ and gate count $G$ of a single time step read
\begin{align}
    D = (d + 1) \, m, \qquad \qquad G = (d+1) \, m  N - bdm = (N-b)D + bm,
    \label{eq:gates}
\end{align}

\noindent where $d$ denotes the spatial dimension and $b$ is 0 for periodic or 1 for open boundary conditions. We assume an optimistic gate fidelity of $p_g = 99.9\%$ \cite{Kjaergaard2020}, so that our results should remain relevant for future years. 

The improvement in $\mathcal{F}_{var}$ and hence in $\mathcal{F}_{NISQ}$ strongly depends on which single time step size $\tau$ is chosen. For the Trotter-Suzuki approximation, a small ratio of $\tau$ and $m$ is favored to keep the error small (cf. (\ref{eq:Ham})). This can be done by increasing $m$ and paying the price of using more gates. Figures \hyperref[fig:Vary_N]{\ref*{fig:Vary_N} b, e} show that for small $\tau$ the gate infidelity dominates the error, whereas for large $\tau$, the algorithmic error takes over. For the one-dimensional case, Trotter-Suzuki error becomes important at around $\tau = 1.5$ for $m = 3$ and at $\tau = 2.0$ for $m = 6$. The variational algorithm has the latest turning point, where the algorithmic error becomes dominant, around $\tau = 2.5$. This turning point is earlier (around $\tau \approx 1.5$) in the two-dimensional model.

If we choose the single time step at this turning point, we find that the variational sequence increases the lower infidelity bound due to gate infidelities only marginally, while the Trotter-Suzuki decompositions introduce a larger additional error.
\begin{figure}
    \centering
    \includegraphics[width=0.975\textwidth]{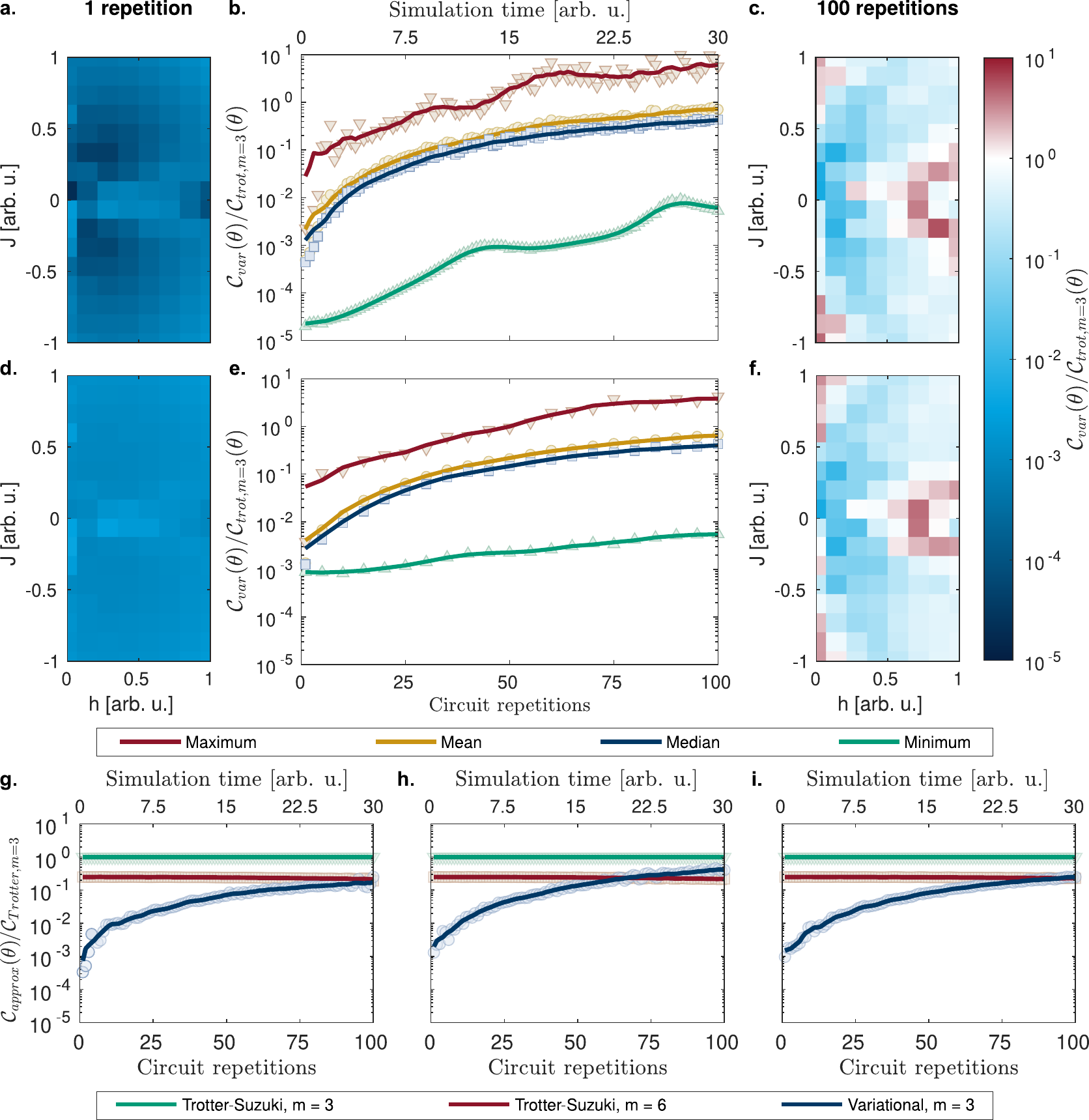}
    \caption{ \textbf{ Improvement of variational vs. Trotter algorithm for a sector of the $J_z - h_x$ parameter space and multiple repetitions.} The parameters are scanned through the values $J_z \in [-1, 1], h_x \in [0, 1]$ and $\tau=0.3$ is fixed. Solid lines denote a moving mean through the simulated data (data points). Using parameters optimized on a 6 qubit chain both, the 6 qubit (\textbf{a. - c.}) and 15 qubit (\textbf{d. - f.}) systems, show similar scaling behaviour in the simulation time. A single time step shows an improvement of cost between $10^{3}$ and $10^{5}$ throughout the scanned values for $J_z$ and $h_x$. Even after 100 repetions for most $J_z$ and $h_x$, the  variational algorithm beats the Trotter-Suzuki algorithm. Only where the Trotter-Suzuki decomposition is nearly exact (when $h_x$ or $J_z$ are almost 0), Trotter-Suzuki overtakes before 100 repetitions. In the worst case scenario, the variational algorithm gets overtaken at around 50 time steps. Figure \ref{fig:NISQ} shows that after 50 repetitions, however, the circuit infidelity is typically dominated by gate imperfections. The scaling in simulation time is demonstrated for the cases of open boundaries on a one-dimensional system with 15 qubits (\textbf{g.}), as well as a 2D a system with periodic (\textbf{h.}) and half-open boundaries (\textbf{i.}) of size $3 \times 5$. Here, the interactions $J_z = 1.0$ and $h_x = 0.25$ are chosen as an instance of the average scenario.}
    \label{fig:Vary_T}
\end{figure}
For example, with 50 Qubits and periodic boundaries in one dimension the NISQ infidelity $1-\mathcal F_{NISQ} \approx 29 \%$ is increased by only $7 \%$ due to the variational approximation infidelity, whereas $1 - \mathcal F_{NISQ} \approx 57 \%$ for the best ($m=6$) Trotter-Suzuki ansatz. Both, a higher gate count and approximation infidelity lead to this deterioration. The standard deviation $\sigma$ of the linear $\mathcal F_{approx}$ extrapolations in Fig \hyperref[fig:Vary_N]{2 c.} and \hyperref[fig:Vary_N]{f.} are small compared to $\mathcal{F}_{approx}$: $\sigma = 0.19 \% \cdot \mathcal{F}_{approx}$ (1D) and $\sigma = 0.03 \% \cdot \mathcal{F}_{approx}$ (2D). Therefore, the fits in Figure \ref{fig:Vary_N} yield a faithful extrapolation in system size.

\subsection{Multiple Time Steps}
Since we determine $\Uvar$ via a state agnostic optimization, it can be applied repeatedly to the time evolved state. In this way larger simulation times $T$ can be covered. Figure \ref{fig:Vary_T} show this for a 6- (\hyperref[fig:Vary_T]{a. - c.}) and a 15-qubit system (\hyperref[fig:Vary_T]{d. - f.}) both using the parameters optimized on 6 qubits. The variational sequence is optimized for a single time step, and improvements in the cost function decrease for later times. Despite this, the variational algorithm remains better than the Trotter-Suzuki benchmark even after one hundred repetitions for most values of the Hamiltonian coefficients $J_z$ and $h_x$. However, when one of these coefficients is small the Trotter error decreases, which can lead to Trotter-Suzuki overtaking the variational algorithm in certain regions of parameter space. Indeed, the Trotter-Suzuki approximation becomes exact in the limit that either $J_z$ or $h_x$ goes to zero. In extreme cases, the accuracy of the Trotter-Suzuki sequence exceeds that of the variational algorithm after 50 repetitions. However, as we shall show later, at 50 repetitions the total error is dominated by gate infidelities rather than approximation error.

We obtain a similar behavior of the cost function for the case of $d=2$ and open boundary conditions. This indicates that the open boundary variation of our method can be applied without significant change of performance. In this numerical demonstration, our variational algorithm exhibits even a better approximation fidelity for half-open boundaries than for periodic boundaries, and hence a fully translational invariant system. This further stresses the feasibility of our method for NISQ-inspired open boundary quantum systems.

\subsection{Precision of Time-Evolved Observables}
When computing the quantum dynamics of a system, one is typically interested in the evolution of physical observables.
From (\ref{eq:upper_bound}), it is clear that the cost values only yield an upper bound for the error in predicting observables. Although the variational algorithm has a smaller cost value, Trotter-Suzuki could thus, in principle, describe the dynamics of a specific observable for a specific initial state with greater accuracy. The approximation error in the expectation value of a time-evolved observable hence depends on the precise observable and on the chosen initial state. In this section, we show results for exemplary observable dynamics and an initial state.

Note that the gluing procedure is not sensitive to entanglement of the input state, since the time evolution operator is agnostic of the initial state. States of high entanglement (or strongly entangling dynamics) are particularly interesting, as the Trotter error of localized physics has been found to be independent of simulation time \cite{Heyleaau8342}.

\begin{figure}
    \centering
    \includegraphics[width=0.975\textwidth]{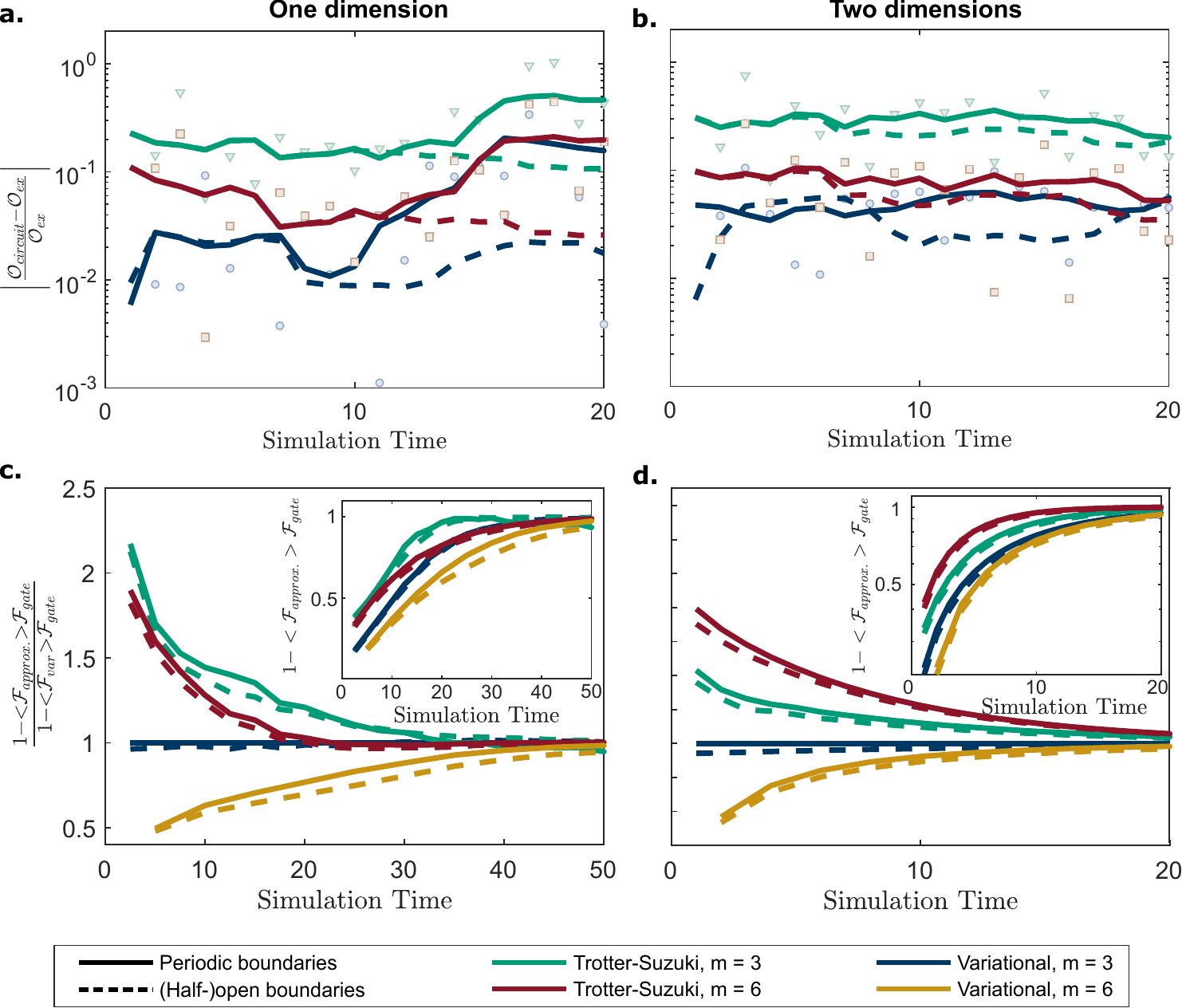}
    \caption{ \textbf{a. - b. Error of the two-point correlation} $\mathcal{O} = <\psi|Z_8 Z_9|\psi>$ between qubit 8 and 9 of a 15 qubit system ($3\times 5$ for $d=2$) for multiple time steps starting with the initial state $\psi = \ket{+}^{\otimes 15}$. The interaction coefficients and the single time step read $J_z, = 1.0$, $h_x= 0.25$ and $\tau = 1.0$. Solid lines denote a moving mean through the simulated data (data points). Although the variational sequence quickly converges to the same error as Trotter with $m=6$, it will be less affected by gate infidelities. \textbf{c. - d. Infidelity including imperfect gates} for multiple time steps with single time step $\tau = 2.5$ ($\tau = 1.0$ for $d=2$). The infidelities have been normalized to the infidelity value of the variational sequence optimized on a single time step with $m=3$ layers. To further reduce the infidelity for longer times, it pays off to optimize a single time step of $2\tau$ including $m=6$ layers and hence with double the number of parameters. Both variational algorithms beat the Trotter-Suzuki benchmarks for up to 8 repetitions. Afterwards every approach is dominated by noise.}
    \label{fig:NISQ}
\end{figure}

In Figure \ref{fig:NISQ}, we illustrate the relative error on two-point correlation of two neighboring spins on a 15 qubit system ($3 \times 5$ for $d=2$) far away from the boundary, $\mathcal{O} = \braket{\psi(t)|Z_8 Z_9|\psi(t)}$ ($\mathcal{O} = \braket{\psi(t)|Z_{(2,3)} Z_{(2,4)}|\psi(t)}$ for $d=2$). We find that for the initial state $\ket{\psi(0)} = \ket{+}^{\otimes N}$, the two-point correlation $\mathcal{O}$ predicted by $\Uvar$ lies closer to the exact dynamics than any of the Trotter-Suzuki benchmarks.

Taking a product state as initial state is particularly interesting, as it is accessible in experiment, yet also yields an entangled state after a few repetitions. Although the $m=3$ variational circuit and the $m=6$ Trotter-Suzuki benchmark achieve similar approximation error, the latter contains twice as many gates and thus suffers from a squared gate fidelity.

These imperfections are illustrated in Figures \hyperref[fig:NISQ]{\ref*{fig:NISQ} c. - d.}, where we show the overall infidelity of (\ref{eq:cost_infid}), for up to 20 repetitions of the single time step $\tau=1.0$. Besides the variational sequence beating both Trotter-Suzuki benchmarks, the overall infidelity is minimized even more when optimizing for a larger time step $2 \tau$ using $m=6$ layers. Despite a significant noise floor set by the gate infidelities, we achieve an improvement of infidelity by a factor between 3 ($d=2$) and 4 ($d=1$) for a single repetition. Since the best Trotter-Suzuki decomposition starts with an infidelity of $44.5 \%$, this shows that the simulation time $2\tau = 5.0$ cannot yield satisfactory results in the sketched NISQ scenario, while the variational sequence does (infidelity of $18.8\%$). Since also twice as many parameters are involved, it is expected that the variational sequence optimized on longer times performs better. This comes at the cost of a more difficult optimization task in the classical pre-processing step. The trade-off between more time spent once in classical pre-processing against a permanently better quantum circuit is very favourable on NISQ devices. Generalizing this to longer simulation times $k\tau > 5.0$ and more parameters $m>6$, however, does not further improve the long time performance and thus is omitted in Figure \ref{fig:NISQ}. 

All the numerical experiments support the maxim to keep the number of repetitions low while using as large time steps $\tau$ as possible. To do this, an increase of the Trotter number for the variational single time step $m$ might be worth evaluating (cf. Figure \ref{fig:NISQ}). For small $\tau$ (as $\tau = 0.3$ in Figures \ref{fig:Vary_N} and \ref{fig:Vary_T}, for instance), raising the Trotter number $m$ does not significantly improve the cost values. Therefore, the data points for $m>3$ are omitted. 

Also remarkably, the (half-)open boundary systems yield even smaller infidelity values than their periodic counterparts. Besides supporting the statement that open boundaries can be glued to get NISQ-friendly systems, this phenomenon can be explained, since in open boundary systems fewer gates contribute to both approximation and gate error.

\section{Conclusion}
In this work, we demonstrated how classical pre-processing can be leveraged to determine a variational representation of time evolution operator of a translation symmetric quantum system that reliably exceeds comparable Trotterization in approximation accuracy. The main idea for our variational algorithm consists of classically optimizing a variational time evolution ansatz circuit on a small tractable system and then reusing these optimal parameters for larger, classically intractable systems by gluing gate sequences together. 

These results are particularly interesting due to their applicability to NISQ devices. Most promisingly, it can avoid any sampling cost and limitation associated with variational optimization on quantum hardware. Further, the significant increase in approximation fidelity compared to Trotter-Suzuki with the same gate count for practically relevant system sizes can be taken advantage of to realize the same approximation fidelity at a notably reduced gate count. We found that, for one- and two-dimensional example problems, the observed advantages persist and are in fact enhanced for systems with open boundaries, despite the loss of translational symmetry on the edges. Such open boundary systems are important as they are easier to realize on NISQ hardware.

Our results open up a highly versatile method for variational quantum simulation. Since we have shown that it is sometimes sufficient to maximize the simulation quality locally, possible limitations of this gluing scheme are worth investigating further. We also expect to gain further insights into the scope and performance of our method from investigations of the relations between the range of correlations in simulated quantum states and the size of the system considered in classical optimization. Since the optimization is however state agnostic, depending only on the Hamiltonian of the system, it is expected to perform well in simulating the time evolution of highly-entangled states with long-ranged correlations.

Another way of optimizing quantum simulation algorithms can be achieved by a quantum optimization. The cost function of (\ref{eq:Frob_cost_samp}) can be measured by comparing initial and output state of a circuit which first performs the variational ansatz circuit which propagates forward in time. Afterwards the state is propagated back to its initial state by a high order Trotter-Suzuki approximation. This way, the variational circuit can be optimized to invert the Trotter-Suzuki decomposition with fewer gates and later scaled up in system size. Further investigation is needed to determine which system sizes are suitable for optimization.

Randomization techniques have attracted much attention recently and have been shown to compress Trotter circuits \cite{Campbell19, Childs2019fasterquantum}. It remains to be investigated whether randomized ansätze can be used for variational algorithms like the one presented in this work.

\ack
We thank Petr Zapletal for valuable feedback and discussions and Naeimeh Mohseni for feedback on the manuscript. This work received funding from the European Union's Horizon 2020 research and innovation program under Grant Agreement No. 828826 ``Quromorphic.'', from the German Federal Ministry of Education and Research via the funding program quantum technologies - from basic research to the market under contract number 13N15577 ``MANIQU". It is also part of the Munich Quantum Valley, which is supported by the Bavarian state government with funds from the Hightech Agenda Bayern Plus. TE acknowledges funding from the International Max Planck Research School Physics of Light.

\section*{Data Availability Statement}
The data that support the findings of this study are available upon reasonable request from the authors.

\newpage
\appendix
\section{Higher Order Trotter-Suzuki Decomposition}
\label{app:2OSuzuki}
\begin{figure}[H]
    {
    \centering
    \includegraphics[width=0.975\textwidth]{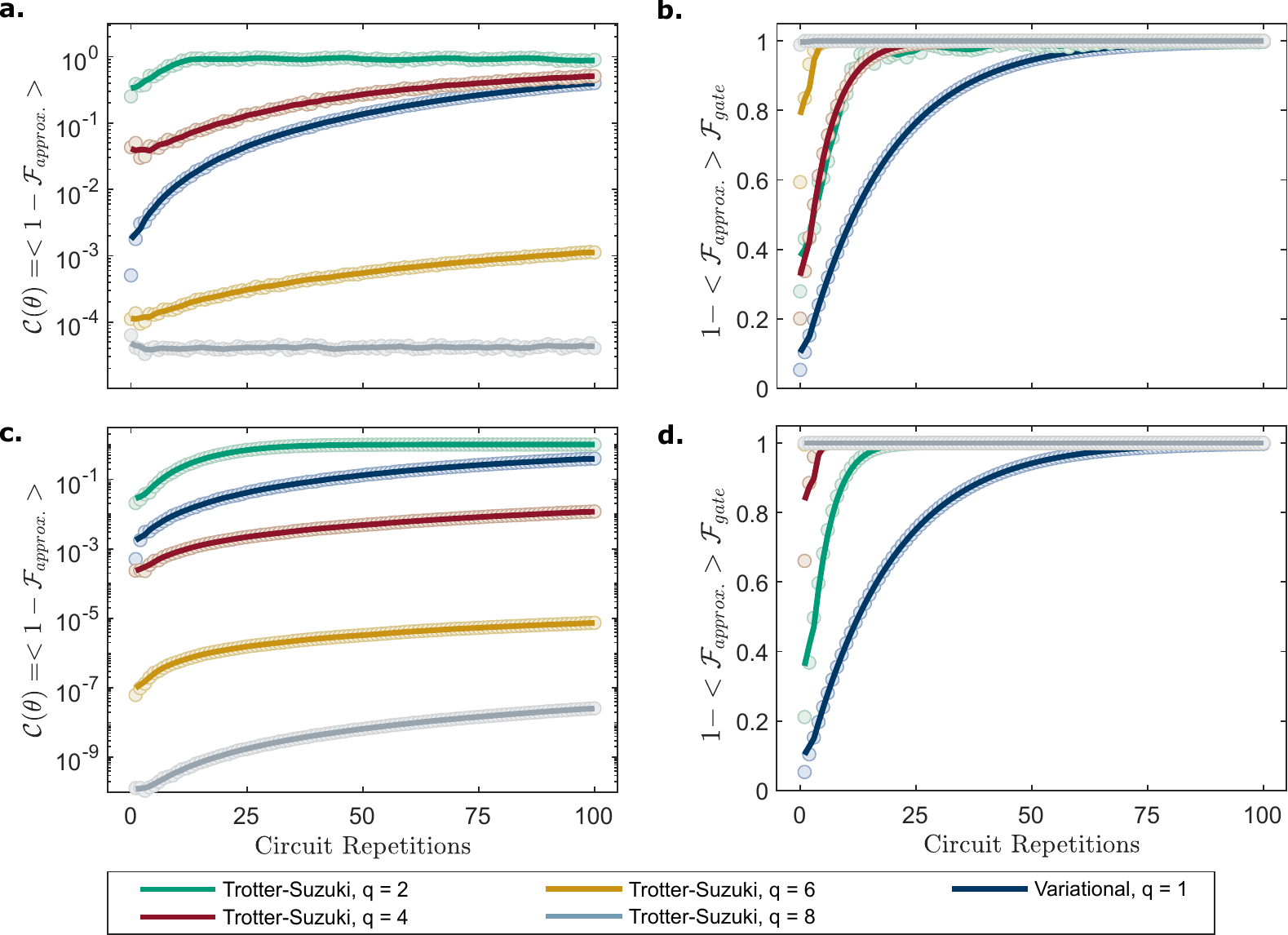}
    }
    \caption{ \textbf{Comparision of approximation and preperation fidelity of variational algorithm and higher order $q$ Trotter-Suzuki simulation} for up to 100 repetitions of time evolution under a \ham with $J_z=1.0, h_x = 0.25, \tau = 1.0$ of a $3 \times 6$ qubit system. In a, b. the Trotter number is $m=1$ and in c, d. $m=2$. The variational algorithm always has $m=3$. Solid lines denote a moving mean through the simulated data (data points). \textbf{a, c.} The variational algorithm beats the $4^\text{th}$ order product formula for $m=1$ while using only $30 \%$ of the gates. The $6^\text{th}$ and $8^\text{th}$ order product formula always beat the variational algorithm in approximation accuracy but need more gates by a factor of 50 and 250. \textbf{b, d.} Taking into account a finite gate fidelity of $p_{g} = 99.9 \%$ together with the algorithmic fidelity $\mathcal{F}_{approx}$ yields the total infidelity of the algorithm. Already after a single time step, the gate fidelity becomes dominant, such that the variational algorithm outruns all higher order product formulas.}
    \label{fig:higher_order}
\end{figure}

A more accurate decomposition as in (\ref{eq:Trotter1}) can be achieved by higher order product formulas. A second order Trotter-Suzuki decomposition is achieved by only reordering the operators in a symmetric way. For a single Trotter step ($m=1$) this reads
\begin{align}
    \Uex &= \left[ \prod_{n, a}^\leftarrow \exp{\left( \frac{- i \tau}{2} c_a H_{n, a} \right)} \prod_{n, a}^\rightarrow \exp{\left( \frac{- i \tau}{2} c_a H_{n, a} \right)} \right] + \mathcal{O}\left( \tau^3 \right) =: S_2(\tau),
    \label{eq:Trotter2}
\end{align}
Furthermore, one can achieve arbitrarily accurate approximations of an exponential $e^{-i\tau H}$ by applying the $q^\text{th}$ order Trotter-Suzuki formula which is defined recursively by \cite{Suzuki_85}
\begin{align}
    S_{q}(\tau) = [S_{q - 2} (\nu_q \tau)]^2 S_{q - 2} ((1 - 4 \nu_q) \tau)[S_{q - 2} (\nu_q \tau)]^2, \quad \text{ for } q > 2
    \label{eq:higher_order_suzuki}
\end{align}
with $\nu_q = \left( 4 - 4^{\frac{1}{q - 1}} \right)^{-1}$. In addition to (\ref{eq:higher_order_suzuki}), one can also pick a Trotter number $m>1$ to raise the accuracy further. Employing a higher $m$ is achieved by substituting $\tau \mapsto \frac{\tau}{m}$ in (\ref{eq:higher_order_suzuki}), but repeating the sequence $m$ times, thus using more gates by a factor of $m$.

We will stick to the $m \in \{1, 2\}$ case here and show that the use of higher order formulas as a benchmark for our method is practically not relevant for NISQ-focused variational optimisation, since the number of required gates, which scales with $5^q$ \cite{Berry06}, will dominate the error for finite gate infidelities. We provide results for $q \in \{2, 4, 6, 8\}$.

The overall circuit fidelity is given by the product of the imperfect gate fidelity $\mathcal{F}_{gate}$ and the fidelity due to the approximate algorithm
\begin{align}
    \mathcal{F}_{var}(\ket{\psi}) = \braket{\psi |U_\text{ex}^\dagger U_\text{var} | \psi}
\end{align}
for an arbitrary initial state $\ket{\psi}$. Analogously, one can define the algorithmic fidelity $\mathcal{F}_{trot}$ of a Trotter-Suzuki algorithm described by the product formula in (\ref{eq:higher_order_suzuki}). We combine the notations $\mathcal{F}_{var}$ and $\mathcal{F}_{trot}$ into $\mathcal{F}_{approx}$. Figure \ref{fig:higher_order} shows the total infidelity $1 - \mathcal{F}_{gate} \cdot \mathcal{F}_{approx}$ for up to 100 time steps. One can see that while second order Trotter-Suzuki would admit fewer gates than the variational algorithm ($m=3$), it suffers from greater algorithmic errors. Higher order Trotter-Suzuki algorithms are already dominated by gate infidelities after a single time step. The variational algorithm hence outruns all of the benchmark algorithms in the presented simulation problem. For this reason, no higher order $q$ Trotter-Suzuki formulas are discussed in the main text. Other product formulas \cite{YOSHIDA1990262, Hatano2005} behave very similar to Trotter in this respect. As Trotter exhaust theoretical complexity lower bounds for higher orders \cite{Childs21}, we only discuss higher order Trotter formulas here.

\section{Smallest Two-dimensional \model Model with Fully Open {Boundary} Conditions}
\label{app:2D_open}
\begin{figure}
    \centering
    \includegraphics[width=0.5\textwidth]{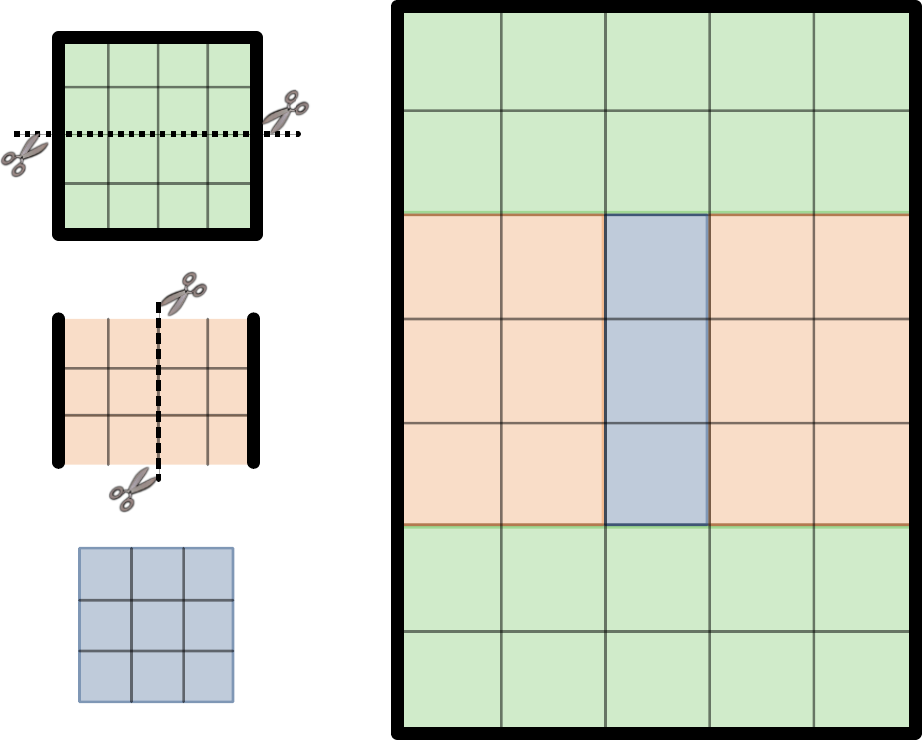}
    \caption{\textbf{ Smallest possible, non-trivially glued configuration of a fully open two-dimensional qubit grid. } The building blocks, on which the optimization is performed, are shown on the left and the fully glued grid on the right and bold lines depict open boundaries. The minimal size of a half-open grid is $3 \times 4$ (orange) and of an open grid $4 \times 4$ (green). The minimal extension of the bulk then consists of three qubits (blue) thus leaving an open $5\times 7$ grid. }
    \label{fig:2d_open}
\end{figure}

We construct the smallest non-trivially glued fully open boundary system in two dimensions. It turns out that, although optimal parameters to use on fully open two-dimensional systems can be found, no measure of accuracy can be numerically evaluated. In  Figure \ref{fig:2d_open} every colored square corresponds to a qubit and thick lines represent open boundaries. As every qubit contributes to the Hamiltonian with a number of $A$ interactions, the squares can be thought of as a set of $A \cdot m$ optimal parameters and the figure depicts how parameters are copied when scaling up. The blue squares all share the same parameters for each gate.

Optimized open boundary systems have to be cut in half to insert bulk qubits inbetween (cf. section \ref{sec:glue}). In two dimensions however, this means the system needs to be cut along both dimensions. Then first, a half-open system is inserted to faithfully keep the open boundary upright (orange blocks in Figure \ref{fig:2d_open}). Finally, completely periodic qubits can be filled in the very bulk of the system (blue blocks), but also the green blocks have added at least one column of qubits to keep everything rectangular. The additional green column copies parameters from the inner most qubit of one half of the boundary (for one column it does not matter whether it copies the parameter from the left or the right). This way, the inner most qubit of each half of the fully open system has to be considered equivalent to a qubit from a half-open system. If we would not choose at least 4 qubits along this open boundary, at least one boundary qubit would coincide with a bulk qubit which is contradictory.
    
If there is a periodic boundary along one dimension nothing has to be cut in half, however, the optimization should treat every qubit such that it has two distinct neighbors along the periodic dimension. This can only be achieved with at least three qubits, as otherwise left and right neighbors would coincide. As a result, the smallest periodic boundary block is of size $3\times 3$, the smallest half-open block is $3 \times 4$ qubits (3 along the periodic boundary and 4 along the open boundary) and the smallest fully open block is of size $4 \times 4$.

Thus, the smallest glued open boundary system in two dimensions is of size $5 \times 7$ consisting of three optimized blocks, a half-open one of size $3 \times (2+2)$, a fully open block of size $5 \times (2+2)$ (the optimized block is of size $(2+2) \times (2+2)$) and a row of 3 qubits in the middle, see Figure \ref{fig:2d_open}. The optimization of the single elements can all be performed on a classical computer, but the cost function for a $5 \times 7$ qubit system can no longer be evaluated, as we exceed our limit of numerical feasibility. The results for a half-open system of size $3 \times (4+K)$ can be evaluated and hence represent a numerical example of gluing open boundaries to two-dimensional spin lattice models.

\section{The XY Model with a Transverse Field}
\label{app:XY}
\begin{figure}
    \centering
    \includegraphics[width=0.95\textwidth]{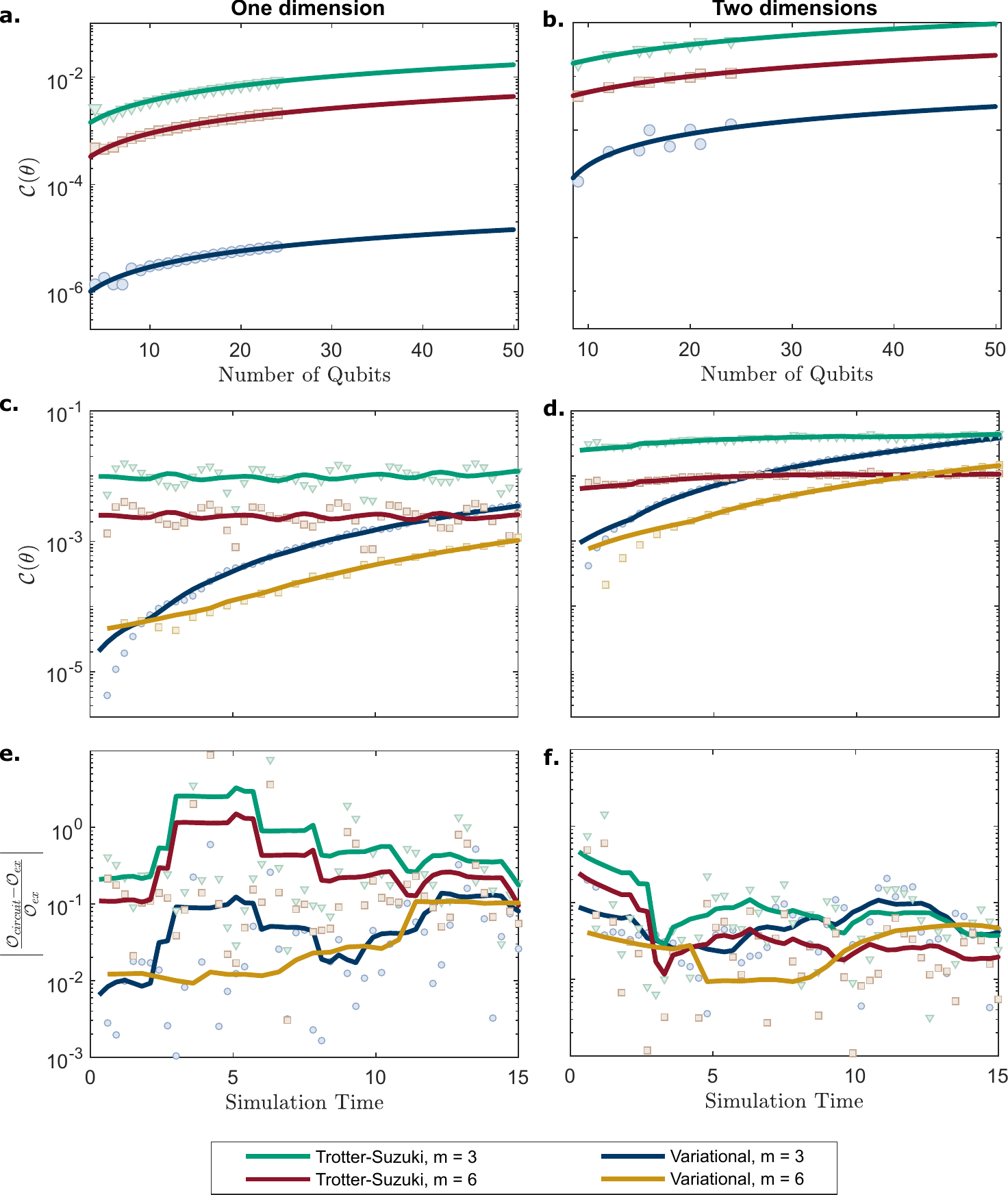}
    \caption{\textbf{ System size and simulation time scaling of variational and Trotter sequence for the XY Model.} In the simulations of several system sizes and for one- and two-dimensional systems with periodic boundaries, the interaction coefficients $J_y=0.5, J_z = 1.0, h_x = 0.25$ and a single time step of $\tau = 0.3$ ($\tau = 0.6$ for Variational $m=6$) were used. As in before, solid lines in a. and b. are linear fits and moving means through the simulated data (data points) in c - f.. \textbf{a. -- b.} Comparison of cost function for a single time step $\tau$ and different system sizes. The optimization was performed on 6 qubits for the one-dimensional and on a $3 \times 3$ grid for the two-dimensional system. The data points represent numerical evaluations of the cost function whereas the solid lines represent a linear fit in the number of qubits. 
    \textbf{c. - d.} Cost values for multiple time steps on a one-dimensional system with 15 qubits and a two-dimensional system of size $3 \times 5$. For the XY model, the choice of $\tau = 0.6$ and $m=6$ yields a better result for scaling in time.
    \textbf{e. - f.} Relative error of the two-point correlation $\mathcal{O} = \braket{Z_8 Z_9}$ for the same systems as in c. - d.. The initial state is chosen to be $\psi = \ket{+}^{\otimes 15}$. Although the cost value is generally smaller in the one-dimensional case (cf. a. - d.), the Trotter-Suzuki sequence approximates the chosen observable better in the two-dimensional case, reducing the improvement by optimization. In \textbf{c. - f.} solid lines denote a moving mean through the simulated data (data points). }
    \label{fig:XY_model}
\end{figure}

Next to the study of the \model Model, in this appendix the $XY$ model with a transverse field (TFXY) is discussed using similar quality measures as in section \ref{sec:numerics}. Coming from the \model Model, we just add a $YY$ interaction to neighboring qubits and get an $XY$ model with an external field. The Hamiltonian reads
\begin{align}
    H_{TFXY} = J_y \sum_{\langle i, j \rangle} Y_i Y_j + J_z \sum_{\langle i, j \rangle} Z_i Z_j + h_x \sum_i X_i.
    \label{eq:YY_Ham}
\end{align}
Since this model includes more non-commuting terms, we expect the Trotter error to be larger, making it another interesting model for NISQ simulation. This claim is supported by Figure \ref{fig:XY_model} which compares cost values of variational and Trotter-Suzuki sequence scaled in system size and in simulation time, as well as errors of observable dynamics. 

Figures \hyperref[fig:XY_model]{\ref*{fig:XY_model} a. - b.} show that the first order Trotter-Suzuki approximation has a larger cost value compared to the TFXY Model with the same interaction coefficients as in section \ref{sec:numerics} except adding a $YY$ interaction with $J_y = 0.5$. As before, the improvement from optimization of a single time step is larger in one-dimensional than two-dimensional systems and also the scaling in system size, resembles the behavior of the \model Model well. The extrapolation of the cost value in systems size keeps up an improvement in cost value of a factor of 1000 (100 for $d=2$) compared to Trotter-Suzuki with Trotter number $m=3$. For a one-dimensional system with 50 qubits, for instance, the Trotter-Suzuki decomposition cost value is extrapolated to around $0.01$ for $m=3$ and $5 \cdot 10^{-3}$ for $m=6$. The variational sequence has a cost value of $10^{-5}$. For a two-dimensional grid of 50 qubits, in total, the cost values are of the order of $10^{-1}$ (Trotter $m=3$), $10^{-2}$ (Trotter $m=6$) and $10^{-3}$ (Variational $m=3$).

The scaling in simulation time is shown in Figures \hyperref[fig:XY_model]{\ref*{fig:XY_model} c. - f.}. The general trend of the plots again resembles the case of the \model Model which indicates a certain model independence of the variational method. The slight increase in approximation error in the $XY$ model however also reduces the number of circuit repetitions which can be performed until the variational sequence falls back to accuracies of Trotter-Suzuki. For an optimization on $m=3$ layers, this happens after around 40 repetitions in the one-dimensional, and 20 repetitions in the two-dimensional case. To push this limit to longer times, an optimization on double the time step $2\tau = 0.6$ and double the layers $m=6$ can be performed. The improvement gained for larger time steps is traded off against a more complex classical optimization algorithm, here (cf. Figure \ref{fig:NISQ} and discussion).

The simulation of observable dynamics shall complete our excursion to the $XY$ model. In Figures \hyperref[fig:XY_model]{\ref*{fig:XY_model} e. - f.}, the relative error of a two-point correlation along 50 repetitions is shown. The initial state $\psi = \ket{+}^{\otimes 15}$ and observable $\braket{Z_8 Z_9}$ are the same as in the analogue discussion for the \model Model. What is different is the single time step $\tau = 0.3$ and that we also compare the variational sequence optimized on $m=6$ layers. Although the cost values of the two different variational sequences show a clear hierarchy for multiple time steps, there is no clear winner identifiable when it comes to the approximation of the observable dynamics. For the two-dimensional system, the $m=6$ Trotter-Suzuki sequence sometimes yields better approximations to the chosen observable, whereas it is beaten in all repetitions in the one-dimensional case. This emphasizes the volatility of observable errors, although also here an advantageous behavior of the variational algorithms is visible. Note that still $m=6$ Trotter-Suzuki uses double the number of gates than the other sequences which needs to be taken into account in the case of imperfect gate fidelities.

\newpage

\vspace{5ex}
\fadedline{0.2}
\vspace{2.5ex}

\bibliographystyle{unsrt}
\bibliography{literature}

\newpage

\end{document}